# Attractors of Parikh mapping iterations

Alexander Yu. Chunikhin

**Palladin Institute of Biochemistry**

**National Academy of Sciences of Ukraine**

alexchunikhin61@gmail.com

https://orcid.org//0000-0001-8935-0338

**Abstract**. Three types of the Parikh mapping are introduced, namely, alphabetic, alphabetic-basis and basis. Explicit expressions for attractors of the k-th order in bases **n** ≥ 8, including countable ones, are found. Properties for the alphabetic, alphabetic-basis and basis Parikh vectors are given at each step of the Parikh mapping. The maximum number of iterations leading to attractors of the k-th order in the basis **n** is determined.

1. Introduction

The Parikh mapping maps a word over an alphabet with *n* letters to an n-dimensional vector whose components give the number of occurrences of the letters in the word [1].

In [2, 3] it was proposed to modify the Parikh mapping using numbers of the alphabet (basis) as separators between the number of occurrences. Moreover, in the Parikh mapping cardinal numbers and separator numbers were used equally (they didn't differ). Researchers [2, 3] found some attractors of the Parikh mappings, but only partially answered the questions they posed. How many singleton-attractors and cycle-attractors exist for a given base *p*? For a given base *p*, are there both singleton- and cycle-attractors? Are there cycles of arbitrary large size [2]? Theorems that substantiate the existence of certain attractors were formulated, but did not directly present them. Some statements about the number of attractors and multiplicities of occurrences of elements are expressed in the form of inequalities [2].

In this work, we propose to abandon separator numbers due to their unreasonable use. Firstly, separating the number of occurrences is meaningless for the Parikh mapping, because only the actual number of i-cardinals matters, and not their order, location, and, moreover, separation by something. Secondly, the participation of separator numbers in subsequent Parikh mapping iterations leads to unjustified mixing of the cardinal distribution of the components in the Parikh vector. This is precisely what, in our opinion, did not allow the authors of [2, 3] to obtain a general expression for attractors and determine the number of iterations required for this.

## 2. Preliminaries

It is proposed to distinguish between the following types of the Parikh mapping: alphabetic, alphabetic-basis and (purely) basis. Let $V = \{a, b, \ldots, l\}$ be some alphabet, $V^*$ - a set of words $w$ over the alphabet V, $w \in V^*$.

**Definition 1.** The *alphabetic Parikh mapping* is a mapping that associates each word $w \in V^*$ with a multiset $[\#_a, \#_b, \ldots, \#_l]$, where $\#_i$ means the number (occurrence) of letters i ($i \in V$) in the word $w$. To each multiset $[\#_a, \#_b, \ldots, \#_l]$ we associate a vector $|\#>_V = (\#_a\ \#_b\ \ldots\ \#_l)^T$, which we call the *alphabetic Parikh vector*.

Then $\mathcal{P}_V(w): w \to |\#>_V$ is the alphabetic Parikh mapping of the word $w$ into the vector $|\#>_V$. Let $P_V$ be a set of all alphabetic Parikh vectors. Then $\mathcal{P}_V: V^* \to P_V$.

**Example 1.** Let $V = \{a, b, c\}$, and $w = baacab$. Then $\mathcal{P}_V(w) = |\#>_V = (3_a\ 2_b\ 1_c)^T$.

Let us introduce the concept of the basis Parikh vector. Let $\mathbf{n} = \{0, 1, 2, \ldots, n-1\}$ be a numerical basis or a numerical alphabet - an ordered set of numbers. It is natural to define the *basis Parikh vector* as $|\#>_n = (\#_0\ \#_1\ \#_2\ \ldots\ \#_{n-1})^T$. Let $P_n$ denote a set of all n-component basis Parikh vectors (Parikh n-vectors).

**Definition 2.** The *alphabetic-basis Parikh mapping* is a mapping that associates each alphabetic Parikh vector $|\#>_V \in P_V$ with the basis Parikh vector from the set $P_n$ ($|\#>_n \in P_n$):
$$\mathcal{P}_{Vn}(|\#>_V): |\#>_V \to |\#>_n \quad \text{or} \quad \mathcal{P}_{Vn}: P_V \to P_n.$$

**Example 2.** Let $\mathbf{n} = \{0, 1, 2, 3\}$ and $|\#>_V = (3_a\ 2_b\ 1_c)^T$. Then $\mathcal{P}_{Vn}(|\#>_V) = \mathcal{P}_{V4}((3_a\ 2_b\ 1_c)^T) = (0_0\ 1_1\ 1_2\ 1_3)^T = (0\ 1\ 1\ 1)^T = |\alpha>_4$.

**Definition 3.** The *basis Parikh mapping* is a mapping that associates each (basis) Parikh n-vector $|\#>_n \in P_n$ with a Parikh n-vector from the same set $|\#>_n \in P_n$: $\mathcal{P}_n(|\#>_n): |\#>_n \to |\#>_n$.

Thus, $\mathcal{P}_n : P_n \to P_n$ is the basis Parikh mapping.

**Example 3.** Let $\mathbf{n} = 4$, $|\#>_n = |\alpha>_4$. Then $\mathcal{P}_n(|\#>_n) = \mathcal{P}_4((0\ 1\ 1\ 1)^T) = (1\ 3\ 0\ 0)^T = |\beta>_4$.

## 3. Attractors in finite bases

Let us study the basis Parikh mappings for convergence. To do this, we continue iterations of the basis Parikh mapping from Example 3.
$$\mathcal{P}_4(|\beta>_4) = \mathcal{P}_4^2(|\alpha>_4) = \mathcal{P}_4((1\ 3\ 0\ 0)^T) = (2\ 1\ 0\ 1)^T = |\gamma>_4,$$

$\mathcal{P}_4(|\gamma>_4) = \mathcal{P}_4^3(|\alpha>_4) = \mathcal{P}_4((2\ 1\ 0\ 1)^T) = (1\ 2\ 1\ 0)^T = |\delta>_4,$ *

$\mathcal{P}_4(|\delta>_4) = \mathcal{P}_4^4(|\alpha>_4) = \mathcal{P}_4((1\ 2\ 1\ 0)^T) = (1\ 2\ 1\ 0)^T = |\delta>_4!$ *

Obviously, there is no point in iterating further. We come to the concept of the *attractor* of the Parikh mappings.

**Definition 4**. The *k-th order attractor* of the Parikh mapping in the basis **n** ($A^k_n$) is the Parikh n-vector that, when mapped, transforms into itself after *k* iterations.

$A^1_n = |\#>_n^* \Leftrightarrow \mathcal{P}_n(|\#>_n^*) = |\#>_n^*.$

$A^2_n = (|\#>_n^{**} \leftrightarrow |\#>_n") \Leftrightarrow \mathcal{P}_n(|\#>_n^{**}) = |\#>_n", \mathcal{P}_n(|\#>_n") = |\#>_n^{**}.$

Table 1 shows expressions for the k-th order attractors of the Parikh mapping in the basis **n**, provided that the multiplicity of the occurrence of any element in the alphabetic Parikh vector is less than n: $\forall i, \#_i < n$.

**Table 1**. Values of the k-th order attractors of the Parikh mapping in the basis **n**

| **n** | $A^1_n$ | $A^2_n$ | $A^3_n$ |
|---|---|---|---|
| 2 | - | - | - |
| 3 | - | - | - |
| 4 | 1210<br>2020 | - | - |
| 5 | 21200 | - | - |
| 6 | - | 311100<br>230100 | - |
| 7 | 3211000 | - | 4110100<br>3300100<br>4102000 |
| 8 | 42101000 | 43000100<br>51011000 | - |
| 9 | 521001000 | 530000100<br>610101000 | - |
| 10 | 6210001000 | 6300000100<br>7101001000 | - |
| 11 | 72100001000 | 73000000100<br>81010001000 | - |
| 12 | 821000001000 | 830000000100<br>910100001000 | - |
| ... | ... | ... | ... |
| $\mathbb{N}$ | $\mathbb{N}_0 2_1 1_2 1_\mathbb{N} 0_\omega$ | $\mathbb{N}_0 3_1 1_\mathbb{N} 0_\omega$<br>$\mathbb{N}_0 1_1 1_3 1_\mathbb{N} 0_\omega$ | - |

The following general formula was inductively obtained for the first-order attractors and one of the components of the second-order attractors $\forall n \geq 8$ (as well as for $A^2_6$ and $A^1_7$):

$$A^k_n = ((n-4)_0 \ (k+1)_1 \ (2-k)_2 \ 0_3 \ldots \ 1_{n+k-5} \ 0_{n+k-4} \ \ldots \ 0_{n-1})^T.$$

In compact notation we get $A^k_n = ((n-4)_0 \ (k+1)_1 \ (2-k)_2 \ 1_{n+k-5} \ 0_\omega)^T$, where $\omega$ means "in all other positions" or "for all other components of the vector".

## 4. Properties of Parikh n-vectors

For the alphabetic and basis Parikh vectors at each iteration there are certain properties determined by the components of the corresponding vectors.

For the alphabetic Parikh vector obtained as a result of the alphabetic mapping $\mathcal{P}_V(w): w \to |\alpha>_V$, the following holds:

(i) $\Sigma_i \ \alpha_i = |w|$, where $|w|$ is the length of the word $w$, $i = a, b, \ldots, l$ are V-alphabet letters.

For the Parikh n-vector obtained as a result of the Parikh alphabetic-basis mapping $\mathcal{P}_{Vn}(|\alpha>_V) = |\beta>_n$, the following holds:

(ii) $\Sigma_i \ \beta_i = |v|$ – the number of different letters from the alphabet V in the word $w$, $i = 0, \ldots, n-1$;
(iii) $\Sigma_i \ (\beta_i \cdot i) = |w|$.

For the Parikh n-vector obtained as a result of the first basis Parikh mapping $\mathcal{P}_n(|\beta>_n) = |\gamma>_n$, the following holds:

(iv) $\Sigma_i \ \gamma_i = n$, where n is the dimension of the numerical basis, $i = 0, \ldots, n-1$;
(v) $\Sigma_i \ (\gamma_i \cdot i) = |v|$ - the number of different letters from the alphabet V in the word $w$.

For the Parikh n-vector obtained as a result of the second basis Parikh mapping $\mathcal{P}_n(|\gamma>_n) = |\delta>_n$, the following holds:

(vi) $\Sigma_i \ \gamma_i = n$, $i = 0, \ldots, n-1$;
(vii) $\Sigma_i \ (\gamma_i \cdot i) = n$.

For all subsequent iterations of the basis Parikh mapping properties (vi)–(vii) are preserved. Accordingly, the latter are also valid for attractors.

## 5. Attractors in the countable basis

A set of words V* over the alphabet V is in principle countable, so it makes sense to find an expression for the attractors in the countable basis. Let us denote the cardinal of the countable set by $\mathbb{N}$. Since $\mathbb{N} - 4 = \mathbb{N}$ and $\mathbb{N} + k - 5 = \mathbb{N}$, then the previously obtained expression for the attractor in a finite basis

$$A^k_n = ((n-4)_0 \ (k+1)_1 \ (2-k)_2 \ 1_{n+k-5} \ 0_\omega)^T$$

for the case of the countable basis takes the form of:

$$A^k_n = (\mathbb{N}_0 \ (k+1)_1 \ (2-k)_2 \ 1_\mathbb{N} \ 0_\omega)^T$$

or

$$A^1_\mathbb{N} = (\mathbb{N}_0 \ 2_1 \ 1_2 \ 1_\mathbb{N} \ 0_\omega)^T,$$
$$A^2_\mathbb{N} = (\mathbb{N}_0 3_1 1_\mathbb{N} 0_\omega)^T \leftrightarrow (\mathbb{N}_0 1_1 1_3 1_\mathbb{N} 0_\omega)^T.$$

The correctness of the expressions for the Parikh mapping attractors in the countable basis is verified directly by performing the basis mapping $\mathcal{P}_n(A^k_\mathbb{N})$.

## 6. On the reachability of attractors

The presence of the Parikh mapping attractors necessarily raises the question of the number of iterations required to achieve them. The solution to this question is impossible without introducing the concept of the inverse Parikh mapping.

**Definition 5**. The *inverse basis Parikh mapping* $\mathcal{P}_n^{-1}$ is a multivalued mapping that associates each vector $|\#>_n \in P_n$ with a subset of vectors $S_n \subset P_n$ such that for any $|\#>_n \in S_n$ we have $\mathcal{P}_n(|\#>_n) = |\#>_n$. Thus, $\mathcal{P}_n^{-1}: |\#>_n \to S_n$.

**Lemma.** Any Parikh n-vector $|\#>_n$, in which $\Sigma_i \#_i = n$, $i = 0,\ldots, n-1$, is a *generating vector*, that is, it provides the possibility of at least one inverse basis Parikh mapping $\mathcal{P}_n^{-1}$.

Thus, finding the maximum number of the Parikh mapping iterations leading to the k-th order attractor in the basis **n** requires the following algorithm to be executed.
1. For a given basis **n**, choose an attractor of the k-th order and assume it (or one of its vectors) to be generating.
2. Perform the inverse basis Parikh mapping $\mathcal{P}_n^{-1}$.
3. In the resulting set of the basis n-vectors, we find those for which the following conditions are satisfied: $\Sigma_i \#_i = n$; $\Sigma_i (\#_i \cdot i) = n$, $i = 0,\ldots, n-1$;
4. We consider these n-vectors to be generating and for them we perform the inverse Parikh basis map $\mathcal{P}_n^{-1}$.

5. We continue $\mathcal{P}_n^{-1}$ iterations until there are no n-vectors left in the subset $S_n$ that satisfy the condition $\Sigma_i$ (#$_i$·i) = n, i = 0,… , n-1.

6. In this subset, select a vector(s) that satisfies the condition $\Sigma_i$ #$_i$ = n, i = 0,… , n-1. We assume that it is generating according to the lemma.

7. We add 2 to the number of the last iteration and get the number of iterations required to reach the attractor both from the alphabetic Parikh vector and from a certain number or the set of numbers in the digital alphabet V = {0, 1, ..., n-1} if Parikh mappings of numbers or multisets are considered. In this case, the alphabetic and the alphabetic-basic Parikh mappings coincide.

If we add 3 to the number of the last iteration, we get the number of iterations required to reach the attractor from some word $w \in V^*$.

8. The maximum number of iterations in the chain of inverse Parikh mappings determines the value of the *reachability rate* of the k-th order attractor in the n-basis.

**Table 2**. Attractor reachability rates for n ≤ 11 from alphabetic Parikh vectors

| n | k | 1 | 2 | 3 |
|---|---|---|---|---|
| 4 |   | 4 | - | - |
| 5 |   | 5 | - | - |
| 6 |   | - | 5 | - |
| 7 |   | 3 | - | 5 |
| 8 |   | 4 | 7 | - |
| 9 |   | 5 | 8 | - |
| 10 |  | 5 | 8 | - |
| 11 |  | 6 | 7 | - |

**Theorem**. (On the convergence of Parikh mapping iterations) In any finite basis (n ≥ 4) the sequence of Parikh mappings of an arbitrary word *w* from a set of words V* over the alphabet V converges to the attractor $A^k_n$ of the first or second order (first or third for n = 7) in a finite number of iterations.

The question of reachability in the case of the countable basis remains open, although it can be assumed that it is countable.

## 7. Conclusion

Over the past 20 years, there has been a significant reduction in the number of publications in the field of Parikh mapping iterations and Parikh vectors. Parikh vectors gave way to Parikh matrices, and the iterative approach did not find its further advancement precisely because of, as it seems to us, the methodological mistake in the approach [2, 3] and, as a consequence, the lack of significant results for applications.

In our opinion, the potential of this area is far from exhausted, both in the theoretical sense and in the field of applications. Therefore, in this work we try to reform the iterative approach to the Parikh mapping. It is proposed to distinguish between three types of the Parikh mapping, and also expressions for the attractors of the basis Parikh mapping for an arbitrary basis, including the countable one, are obtained.

It should be noted that the reasoning presented in the article is valid for a limited cardinal number of components of any Parikh vector, namely: $\forall i, \#_i < n$. If this is not the case, then preliminary studies show convergence for $n \geq 8$ to the same attractors, though the reachability rates increase by one.

It is necessary to determine reachability rates in the case of the countable basis. And most importantly, isn't its supposed countability just a theoretical limitation for applications?

The next important direction, in our opinion, is the geometrization of the Parikh mapping. It is associated with the transition from the transformation of words to the transformation of geometric objects, in particular, figures. Preliminary research gives hope that this direction may be useful in solving pattern recognition problems.

**References**

1. R.J. Parikh. *On context-free languages*. Journal of the ACM 13 (1966) 570–581. https://doi.org/10.1145/321356.321364.
2. J. Dassow, S. Marcus, Gh. Paun. *Iterative reading of numbers and "black holes"*. Periodica Math. Hungarica 27 (1993) 137–152. https://doi.org/ 10.1007/bf01876638.
3. J. Dassow. *Parikh Mapping and Iteration*. // Multiset Processing, LNCS 2235, Springer-Verlag Berlin Heidelberg (2001), pp. 85–101. https://doi.org/10.1007/3-540-45523-X_5.